\documentclass[aps,prl,notitlepage,twocolumn,superscriptaddress,10pt,floatfix]{revtex4-2}

\makeatletter
\def\@hangfrom@section#1#2#3{\normalsize\@hangfrom{#1#2}#3}
\def\@hangfroms@section#1#2{\normalsize#1#2}
\makeatother

\usepackage{amsmath,amsfonts,amssymb,bm}
\usepackage{graphicx}
\usepackage{enumerate}

\usepackage{theorem}

\usepackage{tikz}

\definecolor{darkblue1}{rgb}{0.18,0.19,0.57}
\usepackage[colorlinks=true, allcolors=darkblue1]{hyperref}

\newcommand{\nc}{\newcommand}

\nc{\ket}[1]{|#1\rangle}
\nc{\bra}[1]{\langle#1|}
\nc{\ketbra}[2]{|#1\rangle\!\langle#2|}
\nc{\braket}[2]{\langle#1|#2\rangle}
\nc{\braoprket}[3]{\langle#1|#2|#3\rangle}
\nc{\opr}[1]{\operatorname{#1}}
\nc{\avg}[1]{\langle#1\rangle}
\nc{\ketbrasame}[1]{|#1\rangle\!\langle#1|}
\nc{\E}{\mathbb{E}}
\nc{\var}{\operatorname{Var}}

\nc{\hk}[1]{\textcolor{violet}{\textbf{[hk: #1]}}}
\nc{\ch}[1]{\textcolor{purple}{\textbf{[ch: #1]}}}
\nc{\ls}[1]{\textcolor{blue}{\textbf{[ls: #1]}}}
\usepackage[normalem]{ulem}
\nc{\sxz}[1]{\textcolor{violet}{sxz:#1}}

\begin{document}
\title{Revealing Entanglement-Growth Mechanisms through the Magic Barrier}
\author{Lv Zhang}
\affiliation{Beijing National Laboratory for Condensed Matter Physics,
Institute of Physics, Chinese Academy of Sciences, Beijing 100190, China}
\affiliation{School of Physical Sciences, University of Chinese Academy of Sciences, Beijing 100049, China}
\author{Shi-Xin Zhang}
\email{shixinzhang@iphy.ac.cn}
\affiliation{Beijing National Laboratory for Condensed Matter Physics,
Institute of Physics, Chinese Academy of Sciences, Beijing 100190, China}

\author{Heng Fan}
\email{hfan@iphy.ac.cn}
\affiliation{Beijing National Laboratory for Condensed Matter Physics,
Institute of Physics, Chinese Academy of Sciences, Beijing 100190, China}
\affiliation{School of Physical Sciences, University of Chinese Academy of Sciences, Beijing 100049, China}
\affiliation{Beijing Key Laboratory of Advanced Quantum Technology,
Beijing Academy of Quantum Information Sciences, Beijing 100193, China}
\affiliation{Hefei National Laboratory, Hefei 230088, China}
\affiliation{Songshan Lake Materials Laboratory, Dongguan, Guangdong 523808, China}
\author{Shuo Liu}
\email{sl6097@princeton.edu}
\affiliation{Department of Physics, Princeton University, Princeton, New Jersey 08544, USA}

\date{\today}

\begin{abstract}
Quantum entanglement and magic are complementary resources underlying quantum computational advantage, yet their dynamical relation in many-body systems remains poorly understood.  In this Letter, we show that the mechanism of bipartite entanglement growth is encoded in the relative timescale between the entropy-growth-rate peak and the magic barrier, defined as the transient peak of the anti-flatness of the entanglement spectrum.  When entanglement is locally built, the same microscopic process increases the entropy and reshapes the Schmidt spectrum, so the magic-barrier peak occurs in the time window of maximal entropy growth. When entanglement is mainly transported or redistributed, entropy can grow before appreciable spectral non-flatness is generated, naturally separating the two peak times. We demonstrate this distinction in the random-field XXZ chain: the two peaks remain strongly correlated in the thermal regime, while their separation grows systematically across the thermal--MBL crossover. We further validate this theoretical framework by employing Bell-pair initial states alongside a tunable SWAP--Haar random circuit. Our results reveal an intrinsic dynamical connection between entanglement and magic, establishing the magic barrier as a powerful spectral diagnostic of how quantum information is generated, transported, and reshaped. 
\end{abstract}

\maketitle

\textit{Introduction.---}
Quantum entanglement is a defining feature of many-body quantum states
and a central probe of nonequilibrium dynamics. The growth of bipartite
entanglement reveals how quantum information spreads~\cite{doi:10.1126/science.aaf6725,Calabrese_2005,PhysRevLett.111.127205,PhysRevX.7.031016,Eisert2015}
and provides a standard diagnostic of distinct dynamical regimes. For
example, after a global quench in a thermalizing system, the
entanglement entropy typically grows approximately linearly before
saturating to a volume-law value~\cite{Calabrese_2005,PhysRevLett.111.127205,Calabrese_2007,Mezei2017,PhysRevLett.128.080602}, whereas in a
many-body-localized (MBL) system~\cite{PhysRev.109.1492,BASKO20061126,PhysRevB.75.155111,PhysRevB.82.174411,PhysRevB.90.174202,PhysRevLett.111.127201,ROS2015420,Imbrie2016,PhysRevX.5.031033,PhysRevB.91.081103,PhysRevLett.114.140401,PhysRevB.88.014206,RevModPhys.91.021001,Sierant_2025,PhysRevB.105.224203,zhang2019strongweakmanybodylocalizations, PhysRevLett.133.196302,PhysRevB.110.184209,Bauer_2013,PhysRevLett.113.107204,PhysRevLett.119.206602,PhysRevLett.121.206601,PhysRevB.107.024204,LIU20253991} it grows much more slowly, often
logarithmically in
time~\cite{PhysRevLett.109.017202,PhysRevLett.110.260601,PhysRevB.90.174202,PhysRevLett.111.127201,RevModPhys.91.021001,Sierant_2025,PhysRevB.96.020406,PhysRevB.95.024202,PhysRevB.104.214202,xu2026entanglementgrowthstructuredinitial}. Beyond its role as a diagnostic, entanglement is also a key resource for
quantum information processing: highly entangled states are generally
difficult to represent classically, and entanglement is a basic
ingredient in many routes to quantum computational advantage~\cite{PhysRevLett.91.147902,10.1098/rspa.2002.1097,PhysRevLett.86.5188,PhysRevLett.86.910}. Yet entanglement alone does not fully
characterize the complexity or computational power of a many-body
quantum state.

\begin{figure}[!t]
    \centering
    \includegraphics[width=0.92\columnwidth]{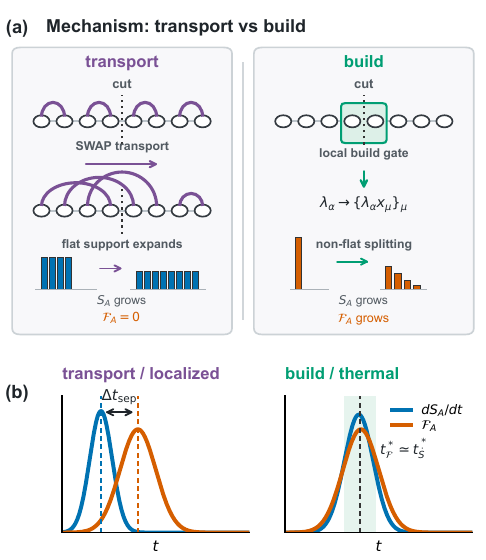}
    \caption{Schematic of build- and transport-dominated entanglement growth.
(a) Transport redistributes pre-existing entanglement across the cut and can
increase $S_A$ through an approximately flat Schmidt block, producing little
or no anti-flatness.  Local build processes instead split Schmidt sectors
nonuniformly, generating ${\cal F}_A$ while increasing $S_A$.
(b) Consequently, transport-dominated dynamics can separate the peak of
$\dot S_A$ from the magic-barrier peak of ${\cal F}_A$, whereas
build-dominated dynamics keeps the two peaks in the same time window.}
    \label{fig:af-schematic}
\end{figure}

Stabilizer states and Clifford circuits provide the clearest example:
they can possess extensive entanglement while remaining efficiently
classically simulable by the Gottesman--Knill
theorem~\cite{gottesman1998heisenbergrepresentationquantumcomputers,PhysRevA.70.052328,PhysRevA.68.042318,PhysRevA.73.022334,PhysRevA.72.022340}. This observation has elevated quantum magic, or nonstabilizerness, to a complementary resource: it quantifies the departure of a quantum state
from the stabilizer manifold and captures the genuinely non-Clifford
content required for universal quantum
computation~\cite{PhysRevA.71.022316,Howard2014,Veitch_2014,PhysRevLett.128.050402,Haug2023stabilizerentropies,PRXQuantum.3.020333,wei2026longrangenonstabilizernessquantumcodes}.
Magic has since been used as a probe of quantum
criticality~\cite{PRXQuantum.4.040317,PhysRevB.103.075145,Tarabunga2024criticalbehaviorsof,pyzr-jmvw,ylsz-dm3y,zhang2026extensivelongrangemagicnonabelian,PhysRevB.111.085144,torre2026nonlocalmagicentanglementspectrum} and has inspired classical
algorithms for quantum many-body problems~\cite{Sun2025stabilizerground,5qr5-7jkz, PhysRevLett.133.190402, PhysRevLett.134.150404,PhysRevB.111.085121}. Its nonequilibrium dynamics has also attracted increasing attention,
with studies in random quantum circuits~\cite{Turkeshi2025,79vj-nx6r,Zhang2026, xiao2026diffusivedynamicsnonstabilizerness, xiao2026exponentiallyacceleratedsamplingpauli,huang2026fastexactapproachstabilizer,Sierant2026computingquantum,feng2026quantumresourcelocalizabilitytransitions}, thermalizing systems~\cite{PhysRevA.108.042407,y9r6-dx7p,1jzy-sk9r}, and MBL systems~\cite{xfp5-hhs4,195d-r5j3}.

Entanglement and magic characterize distinct facets of quantum resources, yet their possible connection in nonequilibrium dynamics remains poorly understood.
The anti-flatness of the entanglement spectrum provides a natural bridge between them: it captures the variance of Schmidt eigenvalues and rigorously lower-bounds nonlocal quantum magic~\cite{PhysRevA.109.L040401,z3vr-w5c5,wtq7-qy11,hvft-hk9p,jasser2026journeyflatlanddoesantiflatness,y9r6-dx7p,PhysRevA.108.042408}.
A numerical correlation between the time of maximal entanglement growth, $t_{\dot S}^*$, and the magic-barrier time, $t_{\cal F}^*$, defined as the peak time of anti-flatness, has been observed in a thermalizing lattice-gauge-theory setting~\cite{wtq7-qy11}.
However, several key questions remain open: how are these two peaks related in more general settings, including nonthermal systems, and, more importantly, what microscopic physical mechanism underlies their correlation?

To answer these questions, we first study the random-field XXZ
chain~\cite{PhysRevB.82.174411,PhysRevB.77.064426,PhysRevLett.109.017202,PhysRevB.91.081103,PhysRevB.105.174205},
where increasing the disorder strength drives the system from a thermal
regime toward an MBL regime.  Starting from random product states, we
analyze the global-quench dynamics of the half-chain entanglement
entropy and anti-flatness.  Deep in the thermal regime, the magic-barrier
time $t_{\cal F}^*$ remains strongly correlated with the time
$t_{\dot S}^*$ of maximal entropy growth, consistent with the behavior
reported in Ref.~\cite{wtq7-qy11}.  Away from this weak-disorder regime,
the two peak times progressively separate, and the separation grows
systematically as the disorder strength is increased.

We provide an analytical understanding of this behavior in terms of two
distinct mechanisms of entanglement growth~\cite{zhang2025entanglementgrowthentangledstates}, as illustrated in
Fig.~\ref{fig:af-schematic}.  In a build process, entanglement is
generated locally across the bipartition: the Schmidt support grows and
the Schmidt spectrum is reshaped by the same microscopic dynamics.
Entropy growth and anti-flatness generation are therefore controlled by
the same local process, leading to strongly correlated peak times.  In a
transport process, by contrast, pre-existing entanglement is redistributed
across the bipartition.  A simple example is a Bell pair moved from
within one subsystem to across the entanglement cut: the entanglement
entropy increases, while the associated Schmidt block remains flat and
therefore contributes little or no anti-flatness.  The two peak times can
then naturally separate.  We verify this build-versus-transport picture
using different initial states in the random-field XXZ chain and a
tunable random circuit composed of SWAP gates and Haar random two-qubit
gates.  Together, the relative timescale between entropy growth peak and the magic
barrier provides a dynamical probe of the entanglement-growth mechanism
and deepens our understanding of the interplay between these two
complementary quantum resources.

\textit{Random-field XXZ model and observables.---}
We consider the one-dimensional random-field spin-$1/2$ XXZ
chain~\cite{PhysRevB.82.174411,PhysRevB.77.064426,PhysRevLett.109.017202,PhysRevB.91.081103,PhysRevB.105.174205},
\begin{equation}
H =
J\sum_{i=1}^{L}
\left(
\sigma_i^x\sigma_{i+1}^x+
\sigma_i^y\sigma_{i+1}^y
\right)
+
\Delta\sum_{i=1}^{L}\sigma_i^z\sigma_{i+1}^z
+
\sum_{i=1}^{L}h_i\sigma_i^z ,
\label{eq:xxz}
\end{equation}
with periodic boundary conditions.  Here $\sigma_i^\alpha$
($\alpha=x,y,z$) denotes the Pauli matrix on site $i$.  We set
$J=\Delta=1$, which fixes the energy and time units, and draw the
random fields independently from a uniform distribution,
$h_i\in[-W,W]$.  The disorder strength $W$ tunes the finite-size system
from a thermal regime toward an MBL regime, with a crossover near
$W\simeq 6.2$ for the sizes studied here~\cite{PhysRevB.105.174205}.
Unless stated otherwise, calculations are performed in the half-filling
sector, $\sum_i\sigma_i^z=0$, starting from $\sigma^z$-basis product
states sampled uniformly within this sector.

For a subsystem $A=\{1,\ldots,|A|\}$, we define $\rho_A(t)=\operatorname{Tr}_{\bar A}|\psi(t)\rangle\langle\psi(t)|$ and compute the von Neumann entropy
\begin{equation}
S_A(t)=-\operatorname{Tr}\rho_A(t)\log\rho_A(t).
\label{eq:SA}
\end{equation}
We also compute the moments $P_n(t)=\operatorname{Tr}\rho_A^n(t)$ and
the anti-flatness of the entanglement spectrum,
\begin{equation}
{\cal F}_A(t)
=
P_3(t)-P_2^2(t)
=
\operatorname{Tr}\rho_A^3(t)
-
\left[\operatorname{Tr}\rho_A^2(t)\right]^2 .
\label{eq:af-def-main}
\end{equation}
If $\{\lambda_\alpha\}$ are the Schmidt eigenvalues of $\rho_A(t)$, then
\begin{equation}
{\cal F}_A(t)
=
\sum_\alpha \lambda_\alpha^3
-
\left(\sum_\alpha \lambda_\alpha^2\right)^2
=
\langle \lambda^2\rangle_\lambda
-
\langle \lambda\rangle_\lambda^2 ,
\end{equation}
where
$\langle f(\lambda)\rangle_\lambda\equiv
\sum_\alpha \lambda_\alpha f(\lambda_\alpha)$ denotes an average over
Schmidt eigenvalues sampled with probability $\lambda_\alpha$.  Thus
${\cal F}_A$ is the variance of the sampled Schmidt eigenvalue.  It
follows that ${\cal F}_A\geq0$, with equality precisely when the nonzero
Schmidt spectrum is flat.  Anti-flatness is therefore sensitive to the
shape of the Schmidt spectrum, not merely to the size of its support.
Unless stated otherwise, we take $A$ to be the half chain, $|A|=L/2$.

We compare two characteristic peak times.  The first,
$t_{\dot S}^*$, is the time at which the entanglement growth rate
$\dot S_A(t)\equiv dS_A(t)/dt$ is maximal.  The second,
$t_{\cal F}^*$, is the time of the maximum of anti-flatness ${\cal F}_A(t)$,
which defines the magic-barrier peak.  Their relative timescale is
quantified by
\begin{equation}
    \Delta t_{\rm sep}=t_{\cal F}^*-t_{\dot S}^* .
\label{eq:tsep}
\end{equation}
All observables are averaged over disorder realizations and initial
states, and $\dot S_A(t)$ is extracted from the averaged entropy trace
by finite differences.

\begin{figure}[t]
\centering
\hspace*{-0.03\textwidth}
 \includegraphics[width=1.05\linewidth]{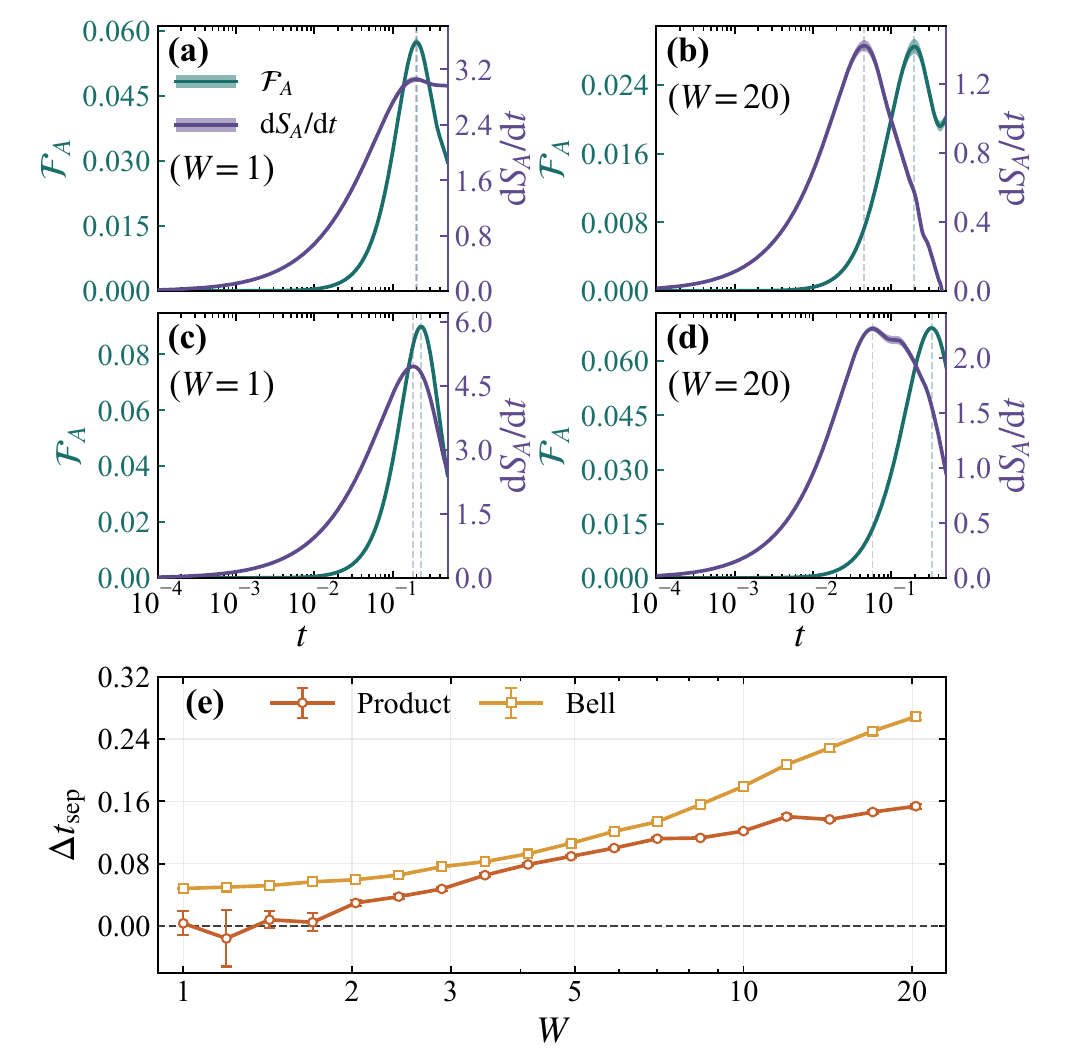}
\caption{
Dynamics of anti-flatness and entropy growth in the random-field XXZ
chain.  The half-chain anti-flatness ${\cal F}_A$ (green) and
entropy-growth rate $\dot S_A$ (purple) are shown for $W=1$ [(a),(c)]
and $W=20$ [(b),(d)] at system size $L=14$.  The upper row uses random
product initial states in the half-filling sector, while the middle row
uses Bell-pair initial states with no Bell pair crossing either
half-chain bipartition boundary.
Vertical dashed lines mark the corresponding peak positions
$t_{\cal F}^*$ and $t_{\dot S}^*$.  All data are averaged over at least
$1500$ disorder realizations.  (e) Peak separation
$\Delta t_{\rm sep}=t_{\cal F}^*-t_{\dot S}^*$ versus disorder strength
$W$ for product and Bell-pair initial states.
}
\label{fig:XXZdynamic}
\end{figure}

\textit{Relative peak timescale across the thermal--MBL crossover.---}
We begin by examining the dynamics of the half-chain von Neumann entropy
$S_A(t)$ and the anti-flatness ${\cal F}_A(t)$ starting from random
product states.  As illustrated in Figs.~\ref{fig:XXZdynamic}(a) and
\ref{fig:XXZdynamic}(b), a transient magic-barrier peak appears over the
range of disorder strengths studied here.  This behavior has a simple
spectral interpretation.  At $t=0$, the product state has a single
nonzero Schmidt eigenvalue across the bipartition, so the Schmidt
spectrum is flat on its support and ${\cal F}_A(0)=0$.  At late times,
the Schmidt support has grown substantially; the corresponding dilution
of the Schmidt eigenvalues suppresses the low moments entering
${\cal F}_A$ (see the Supplemental Material (SM)~\cite{SupplementalMaterials} for an analytical expression of the anti-flatness of random Haar states). The anti-flatness is therefore largest at intermediate
times, after spectral non-flatness has been generated but before it is
strongly diluted by the growth of Schmidt support.

Deep in the thermal regime, exemplified by $W=1$, the magic-barrier peak occurs close to the maximum of the entropy-growth
rate [Fig.~\ref{fig:XXZdynamic}(a)].  Thus \(t_{\cal F}^*\) and
\(t_{\dot S}^*\) are strongly correlated in the thermal regime,
consistent with the behavior reported in Ref.~\cite{wtq7-qy11}.  By
contrast, at strong disorder, exemplified by $W=20$, the two peak times
are clearly separated [Fig.~\ref{fig:XXZdynamic}(b)], indicating that
entropy production and spectral roughening are no longer controlled by a
single local scrambling window.

To chart the crossover between these limits, we extract
$t_{\dot S}^*$ and $t_{\cal F}^*$ as functions of the disorder strength
$W$, as summarized in Fig.~\ref{fig:XXZdynamic}(e).  The separation
$\Delta t_{\rm sep}$ remains small deep in the thermal regime, but
increases systematically as $W$ is increased, revealing a gradual
breakdown of the thermal correlation between entropy growth and the
magic barrier.  This increase is mainly driven by the shift of
$t_{\dot S}^*$ toward earlier times, while $t_{\cal F}^*$ remains
comparatively stable over the same range. Additional details, including numerical results for larger systems and further discussion, are provided in the SM~\cite{SupplementalMaterials}.

\textit{Two mechanisms of entanglement growth.---}
We provide an analytical understanding of the relative peak timescale in
terms of two limiting mechanisms of entanglement growth~\cite{zhang2025entanglementgrowthentangledstates}, illustrated in
Fig.~\ref{fig:af-schematic}.  The first is a local build mechanism, in
which entanglement is generated by local dynamics across the bipartition
cut, e.g., an entangling gate.  At the level of the Schmidt spectrum, a generic build event splits
an existing Schmidt weight into several components,
$\lambda_\alpha\rightarrow \lambda_\alpha x_\mu$, with
$\sum_\mu x_\mu=1$.  This process increases the entropy by the local
Shannon entropy of the splitting and, unless the weights $x_\mu$ are
equal, also creates spectral non-flatness.  Equivalently, the same local
event that expands the Schmidt support also reshapes the Schmidt
weights.  Thus the source of ${\cal F}_A$ and the growth of $S_A$ are
controlled by the same microscopic process, explaining why
$t_{\cal F}^*$ and $t_{\dot S}^*$ are strongly correlated in
build-dominated thermal dynamics.

The second mechanism is transport.  Here pre-existing
entanglement is redistributed across the bipartition rather than locally
created there.  A clean example is obtained by placing Bell pairs inside
the two subsystems and moving one of them across the entanglement cut by
SWAP operations.  The bipartite entropy then increases when the Bell
pair crosses the cut, but the associated Schmidt block remains exactly
flat.  Consequently, this ideal transport process can increase $S_A$
without generating ${\cal F}_A$.  In transport-dominated dynamics, the
entropy-growth rate is controlled by the flux of pre-existing
entanglement through the cut, whereas the magic barrier requires an
additional nonuniform deformation of the Schmidt spectrum by weak build.  An example
with partially entangled dimers is discussed in the
SM~\cite{SupplementalMaterials}.  The two peak times therefore
naturally separate~\cite{SupplementalMaterials}.

This distinction also clarifies why the localized regime is
transport-like in the spectral sense, although it is not a literal SWAP
process.  In the phenomenological $l$-bit description of the MBL
phase~\cite{PhysRevB.90.174202,PhysRevB.91.085425,ROS2015420,PhysRevLett.111.127201,Imbrie2016,PhysRevB.94.144208},
the dynamics is governed by emergent localized conserved quantities with
interactions that decay exponentially with distance.  Entanglement grows
through slow, distance-dependent dephasing across the cut rather than
through rapid local thermal scrambling of Schmidt weights
~\cite{PhysRevLett.109.017202,PhysRevLett.110.260601,PhysRevB.90.174202,PhysRevB.94.144208}.
Different ranges therefore contribute on different dephasing times,
mimicking the sequential arrival or redistribution of entanglement
blocks.  This separates the entropy-growth clock from the clock for
nonuniform spectral deformation, providing a microscopic explanation
for the increasing separation between $t_{\dot S}^*$ and
$t_{\cal F}^*$ in localized dynamics. A related transport-based
interpretation was proposed from a complementary perspective in
Ref.~\cite{zhang2025entanglementgrowthentangledstates}.

\textit{Bell-pair initial states as a stress test.---}
We next test this interpretation within the same random-field XXZ
Hamiltonian by changing only the initial state.  We prepare
nearest-neighbor Bell pairs on dimers contained entirely within either
$A$ or $\bar A$, so that no Bell pair crosses either half-chain
bipartition boundary.  Any
remaining unpaired sites are filled with product spins to keep the state
in the half-filling sector.  The initial bipartite entropy therefore
vanishes, $S_A(0)=0$, although each subsystem contains a reservoir of
short-range entanglement that can later be redistributed across the
half-chain bipartition.

This initial condition enhances the transport-like component of the
dynamics relative to random product states.  We therefore expect a
larger separation between the entropy-growth-rate peak and the
magic-barrier peak at the same disorder strength.  The middle row of
Fig.~\ref{fig:XXZdynamic} shows the corresponding dynamics for $W=1$ and
$W=20$, and Fig.~\ref{fig:XXZdynamic}(e) compares the extracted
$\Delta t_{\rm sep}$ with the product-state result.  The Bell-pair
initial state produces a larger separation over the parameter range, with the enhancement
becoming more pronounced as disorder is increased.  This controlled
change of initial entanglement content supports the interpretation that
the relative peak timescale is sensitive to the balance between local build
and transport-like redistribution~\cite{SupplementalMaterials}.

\begin{figure}[t]
    \centering
    \includegraphics[width=0.47\textwidth]{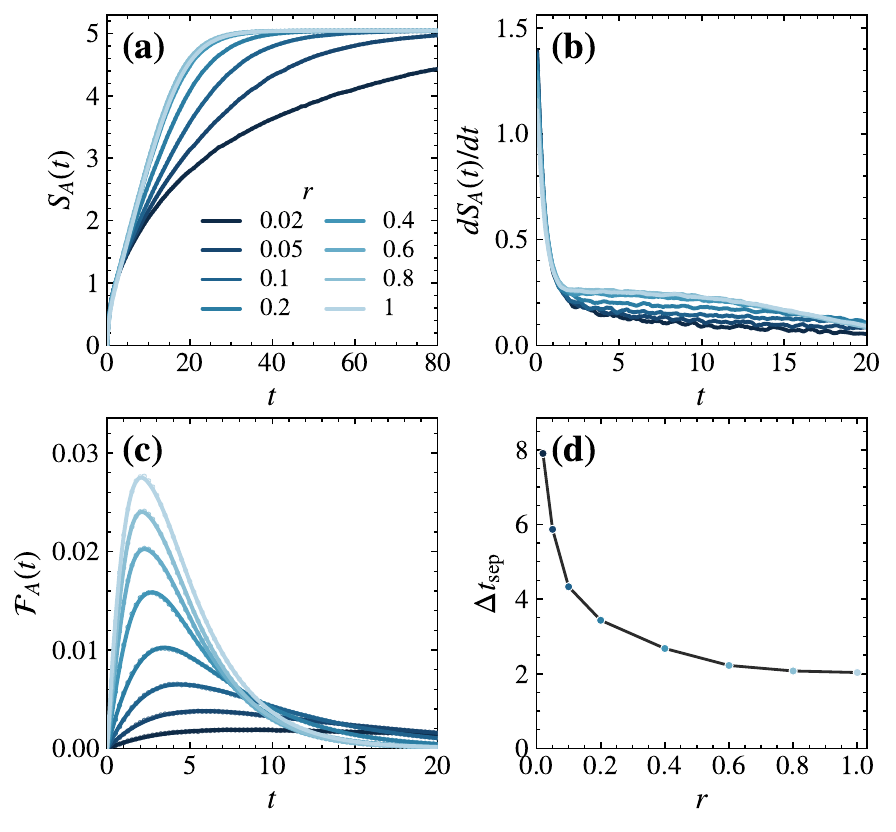}
\caption{
Random-circuit benchmark initialized from Bell pairs.  At each
elementary update, a randomly chosen nearest-neighbor bond is acted on
by a SWAP gate with probability $1-r$ or by a Haar random two-qubit gate
with probability $r$.  The panels show $S_A(t)$, $\dot S_A(t)$,
${\cal F}_A(t)$, and the peak separation $\Delta t_{\rm sep}$ for
$L=16$.  Increasing the Haar-gate fraction strengthens local build
processes and pulls the magic-barrier peak toward the early
entropy-growth window, thereby reducing $\Delta t_{\rm sep}$.
}
    \label{fig:random-circuit-p05}
\end{figure}

\textit{Controlled random circuit benchmark.---}
Finally, we isolate the two mechanisms in a tunable random-circuit
setting.  The system is initialized in the Bell-pair state and evolved
by random nearest-neighbor two-qubit gates.  At each elementary update,
the chosen gate is a SWAP gate with probability $1-r$ and a Haar random
two-qubit gate with probability $r$.  We measure time in circuit sweeps:
one sweep consists of $L$ elementary updates and corresponds to
$\Delta t=1$.  The SWAP gate moves stored Bell-pair entanglement without
locally creating entanglement, whereas the Haar gate locally builds and
scrambles entanglement.  Thus $r$ directly controls the build fraction
of the dynamics.

The results are shown in Fig.~\ref{fig:random-circuit-p05}.  For the
Bell-pair initial state, the entropy growth rate $\dot S_A(t)$ is
maximal at the earliest resolved times for all build fractions shown,
because stored Bell pairs are rapidly transported toward the bipartition
cut.  The anti-flatness peak occurs later: the transported Bell-pair
Schmidt blocks are flat, so a nonzero ${\cal F}_A$ requires
nonuniform spectral deformation generated by Haar updates.  Increasing
$r$ strengthens this local build channel, pulls the anti-flatness peak
toward the early entropy-growth window, and reduces the separation
$\Delta t_{\rm sep}$ (see SM~\cite{SupplementalMaterials} for a detailed discussion).  This controlled benchmark confirms the central
mechanism: entropy growth and the magic barrier separate when the
expansion of Schmidt support is not accompanied by immediate spectral
reshaping.

\textit{Conclusion and outlook.---}
In summary, we have shown that the relative timescale between the
magic-barrier peak and the maximum entropy-growth rate diagnoses the
mechanism of bipartite entanglement growth.  In the thermal regime of
the random-field XXZ chain, the two peak times remain strongly
correlated, consistent with local scrambling processes that both expand
the Schmidt support and reshape the Schmidt spectrum.  As disorder is
increased, this correlation is progressively weakened: the separation
between $t_{\dot S}^*$ and $t_{\cal F}^*$ grows, indicating that entropy
growth and spectral non-flatness are no longer governed by a single
local build clock.  We have provided an analytical interpretation in terms of
two limiting mechanisms.  Local build processes generate entropy and
anti-flatness together, whereas transport-like redistribution or
distance-dependent dephasing can increase entropy before appreciable
nonuniform deformation of the Schmidt spectrum develops.  This picture
is supported by two complementary tests: Bell-pair initial states in the
XXZ chain, which enhance the role of pre-existing internal
entanglement, and a tunable SWAP--Haar random circuit, which directly
interpolates between transport- and build-dominated dynamics.

These results suggest several directions for future work.  First, the
relative timescale between entropy growth and the magic barrier may provide
a useful diagnostic of entanglement mechanisms in broader classes of
nonthermal dynamics, including quantum many-body scars~\cite{Bernien2017,Turner2018,PhysRevB.98.235155,PhysRevB.98.235156,PhysRevLett.122.220603,Serbyn2021,PhysRevLett.132.230403}, discrete time crystals~\cite{PhysRevLett.117.090402,PhysRevLett.118.030401,Zhang2017,PhysRevLett.130.120403}, and systems with Hilbert space fragmentation~\cite{PhysRevX.10.011047,PhysRevB.101.174204,PhysRevX.12.011050,PhysRevLett.133.196301,PhysRevB.110.165109, zhou2026quantumhilbertspacefragmentation}. Such
systems can exhibit slow or constrained information spreading without
conventional thermalization, and it would be natural to ask whether
their entanglement growth is accompanied by local spectral reshaping or
instead proceeds through transport-like redistribution in Hilbert space.
Second, our results point to a broader question: whether other quantum
resources, such as non-Gaussianity~\cite{PhysRevA.78.060303,PhysRevA.97.062337,PhysRevA.98.052350,haug2026practicaltestswitnessesfermionic, w97w-7zny} and coherence~\cite{PhysRevLett.113.140401,RevModPhys.89.041003,PhysRevLett.117.030401,aditya2026coherencedynamicsquantummanybody}, possess dynamical barriers whose timescale is tied to the
growth mechanism of entanglement.  Establishing such connections would
move beyond characterizing resources separately and toward a dynamical
understanding of how different forms of quantum complexity are generated
in many-body systems.

\textit{Acknowledgments.---}
We thank Zhou-Quan Wan for helpful discussions. H.F. and L.Z. were supported by the MOST project (Grant No. 2025YFE0217600), the NSFC (Grants No. U25A6009, No. 92265207, No. 92365301, No. T2121001, No. 92565301), and the QNMP (Grant No. 2021ZD0301800). S.L. was supported by the Gordon and Betty Moore Foundation through Grant No. GBMF8685 toward the Princeton theory program, the Gordon and Betty Moore Foundation’s EPiQS Initiative (Grant No. GBMF11070), the Global Collaborative Network Grant at Princeton University, the Simons Investigator Grant No. 404513, the Princeton Global Network, the NSF-MERSEC (Grant No. MERSEC DMR 2011750), the Simons Collaboration on New Frontiers in Superconductivity (Grant No. SFI-MPS-NFS-00006741-01 and No. SFI-MPS-NFS-00006741-06), the Princeton Catalysis Initiative, and the Schmidt Foundation at Princeton University. S.X.Z was supported by the National Natural Science Foundation of China (No. 12574546), Quantum Science and Technology-National Science and Technology Major Project (No. 2024ZD0301700), and the Chinese Academy of Sciences (No. XDB1680201 and No. YSBR-150).

\let\oldaddcontentsline\addcontentsline
\renewcommand{\addcontentsline}[3]{}%
\bibliography{main}
\let\addcontentsline\oldaddcontentsline
\onecolumngrid

\clearpage
\newpage

\widetext

\begin{center}
\textbf{\large Supplemental Material for \\``Revealing Entanglement-Growth Mechanisms through the Magic Barrier''}
\end{center}

\addtocontents{toc}{\protect\setcounter{tocdepth}{0}}
{
\tableofcontents
}

\renewcommand{\theproposition}{S\arabic{proposition}}
\setcounter{proposition}{0}
\renewcommand{\thedefinition}{S\arabic{definition}}
\setcounter{definition}{0}

\renewcommand{\thefigure}{S\arabic{figure}}
\renewcommand{\theHfigure}{S\arabic{figure}}
\setcounter{figure}{0}
\renewcommand{\theequation}{S\arabic{equation}}
\renewcommand{\theHequation}{S\arabic{equation}}
\setcounter{equation}{0}
\renewcommand{\thesection}{\Roman{section}}

\setcounter{section}{0}
\renewcommand{\thetable}{S\arabic{table}}
\setcounter{table}{0}
\setcounter{secnumdepth}{4}

\section{Spectral diagnostics and build-versus-transport mechanisms}
\label{sec:supp-af-build}

This section provides the theoretical basis for the mechanism discussed
in the main text.  We first show that anti-flatness is a weighted
variance of the Schmidt eigenvalues and then use this spectral
interpretation to distinguish two mechanisms of entanglement
growth: local build and transport-like redistribution.  The remaining
sections provide complementary benchmarks: Sec.~\ref{sec:supp-haar-benchmark}
gives Haar-state baselines, Sec.~\ref{sec:supp-xxz-framework} presents
the random-field XXZ data and peak-extraction conventions, and
Sec.~\ref{sec:supp-random-circuit-benchmark} describes the controlled
SWAP--Haar circuit benchmark.

The central point is that ${\cal F}_A$ diagnoses the \emph{shape} of the
Schmidt spectrum, whereas $S_A$ measures the total \emph{amount} of
bipartite entanglement.  A local build process generically creates new
Schmidt support and non-flat Schmidt weights in the same microscopic
operation.  By contrast, transport-like redistribution can increase
$S_A$ by moving pre-existing correlations across the bipartition without
immediately producing non-flat Schmidt weights.  The relative timescale
between the entropy-growth-rate peak and the anti-flatness peak
therefore distinguishes these two limiting mechanisms.

\subsection{Anti-flatness as a weighted spectral variance}
\label{subsec:supp-af-variance}

For a pure state on $A\cup B$, let
$\rho_A=\operatorname{Tr}_B|\psi\rangle\langle\psi|$ and let
$\{\lambda_\alpha\}$ denote the nonzero eigenvalues of $\rho_A$.  With
\begin{equation}
    P_n=\operatorname{Tr}\rho_A^n=\sum_\alpha \lambda_\alpha^n,
    \qquad
    {\cal F}_A=P_3-P_2^2 ,
    \label{eq:supp-af-def}
\end{equation}
we may sample a Schmidt label $\alpha$ with probability
$\lambda_\alpha$ and define the random variable $X=\lambda_\alpha$.
Then
\begin{equation}
    \mathbb{E}[X]=P_2,\qquad
    \mathbb{E}[X^2]=P_3,
    \qquad
    {\cal F}_A=\operatorname{Var}(X).
    \label{eq:supp-af-variance}
\end{equation}
Thus ${\cal F}_A$ is the variance of the Schmidt eigenvalue sampled from
the Schmidt distribution itself.  It is nonnegative and vanishes
precisely when all nonzero Schmidt weights are equal.  Anti-flatness is
therefore a spectral-shape diagnostic rather than a measure of the
Schmidt support size alone.

This distinction is also transparent in terms of R\'enyi entropies,
$S_n=(1-n)^{-1}\log P_n$:
\begin{equation}
    {\cal F}_A
    =
    e^{-2S_2}
    \left[
        e^{2(S_2-S_3)}-1
    \right] .
    \label{eq:supp-af-renyi}
\end{equation}
The prefactor $e^{-2S_2}=P_2^2$ is a purity, or inverse-support-size,
envelope.  The bracketed factor measures the relative non-flatness
within the occupied Schmidt support.  A large transient value of
${\cal F}_A(t)$ therefore requires two conditions: the spectrum must
become non-flat, and the Schmidt support must not yet be so large that
the low moments are strongly diluted.

Defining
\begin{equation}
    {\cal R}_A=e^{2(S_2-S_3)}-1 ,
\end{equation}
Eq.~\eqref{eq:supp-af-renyi} gives, whenever ${\cal F}_A>0$,
\begin{equation}
    \frac{d}{dt}\log {\cal F}_A
    =
    \frac{d}{dt}\log {\cal R}_A
    -
    2\dot S_2 .
    \label{eq:supp-source-dilution}
\end{equation}
The first term is a spectral-roughening source, while the second term is
the dilution caused by the expanding Schmidt support.  The peak of
${\cal F}_A(t)$ occurs when the roughening rate can no longer compensate
the dilution rate.

\subsection{Local build events}
\label{subsec:supp-build}

The elementary algebra of a local build event can be represented as a
splitting of Schmidt sectors.  As a minimal example, we assume that the
local build event acts with the same splitting pattern on each initially
occupied Schmidt sector, so that the branching weights \(x_\mu\) are
independent of \(\alpha\):
\begin{equation}
    \lambda_\alpha\longrightarrow \lambda_\alpha x_\mu,
    \qquad
    \sum_\mu x_\mu=1 .
    \label{eq:supp-local-splitting}
\end{equation}
The entropy increment is
\begin{equation}
    S'_A-S_A=h[x],
    \qquad
    h[x]=-\sum_\mu x_\mu\log x_\mu ,
    \label{eq:supp-build-entropy}
\end{equation}
while the moments transform as
\begin{equation}
    P'_n=R_nP_n,
    \qquad
    R_n=\sum_\mu x_\mu^n .
    \label{eq:supp-build-moments}
\end{equation}
Combining this with \({\cal F}_A=P_3-P_2^2\) gives
\begin{equation}
    {\cal F}'_A
    =
    R_3{\cal F}_A
    +
    \left(R_3-R_2^2\right)P_2^2 .
    \label{eq:supp-build-source}
\end{equation}
This expression contains two effects.  Since
\(R_3=\sum_\mu x_\mu^3\leq 1\), the first term carries forward the
pre-existing anti-flatness but attenuates it through the expansion of
Schmidt support.  Equivalently,
\begin{equation}
    {\cal F}'_A-{\cal F}_A
    =
    -\left(1-R_3\right){\cal F}_A
    +
    \left(R_3-R_2^2\right)P_2^2 .
    \label{eq:supp-build-source-balance}
\end{equation}
The second term is the local source of anti-flatness generated by the
nonuniformity of the build event.  Its coefficient is
\begin{equation}
    R_3-R_2^2
    =
    \sum_\mu x_\mu^3
    -
    \left(\sum_\mu x_\mu^2\right)^2
    =
    \operatorname{Var}_x(x_\mu),
    \label{eq:supp-build-source-var}
\end{equation}
where the variance is evaluated by sampling \(\mu\) with probability
\(x_\mu\).  This source vanishes only for an equal-weight splitting on
the active local support.

Thus a generic nonuniform build event provides a positive source of
spectral non-flatness while also diluting any anti-flatness already
present.  Starting from a flat or weakly non-flat Schmidt spectrum, this
source is generated in the same microscopic process that increases the
entropy.  In a locally chaotic thermal regime, this common origin
explains why the maximum of \({\cal F}_A(t)\) is expected to occur in
the same early-time window as the maximum of \(\dot S_A(t)\).

\subsection{Transport of stored entanglement}
\label{subsec:supp-transport}

The opposite limit for the entanglement growth mechanism is pure transport.  Suppose entangled dimers are
already present away from the cut and the dynamics only moves them across
the bipartition, as in a SWAP-only dynamics.  For a dimer
\begin{equation}
    |\phi_p\rangle=\sqrt p\,|01\rangle+\sqrt{1-p}\,|10\rangle ,
    \label{eq:supp-dimer}
\end{equation}
let $m(t)$ be the number of dimers that have crossed the cut by time
$t$.  If the crossing dimers are independent, then
\begin{equation}
    S_A(t)=m(t)h(p),
    \qquad
    \dot S_A(t)=h(p)\dot m(t),
    \label{eq:supp-transport-entropy}
\end{equation}
where $h(p)=-p\log p-(1-p)\log(1-p)$.  The entropy-growth rate is
therefore controlled by the instantaneous flux $\dot m(t)$ of stored
entanglement through the cut.

The anti-flatness is controlled by a different quantity: the accumulated
number of crossed dimers.  Since each crossed dimer contributes a
two-level Schmidt spectrum $\{p,1-p\}$, independent crossed dimers give
\begin{equation}
    {\cal F}_A(t)
    =
    \left[p^3+(1-p)^3\right]^{m(t)}
    -
    \left[p^2+(1-p)^2\right]^{2m(t)} .
    \label{eq:supp-transport-af}
\end{equation}
This expression separates two effects.  The entropy-growth rate tracks
the flux of dimers crossing the cut, whereas ${\cal F}_A(t)$ tracks the
spectral shape accumulated after dimers have crossed.  Therefore the
maxima of $\dot S_A(t)$ and ${\cal F}_A(t)$ need not coincide.

The Bell-pair point, $p=1/2$, is the cleanest pure-transport limit.  In
this case each transported Schmidt block is exactly flat, with
$p^3+(1-p)^3=[p^2+(1-p)^2]^2=1/4$, and hence
\begin{equation}
    {\cal F}_A(t)=0
    \qquad
    \text{for all }m(t).
    \label{eq:supp-bell-flat}
\end{equation}
Thus SWAP transport of Bell-pair entanglement can increase $S_A$ but
does not produce an anti-flatness peak.

For partially entangled dimers, $0<p<1$ and $p\neq1/2$, the transported
Schmidt block is already non-flat.  In this case SWAP dynamics can make
the anti-flatness across the cut nonzero, not by generating new spectral
roughness, but by transporting pre-existing non-flat Schmidt weights
into the bipartition.  Writing
$b=p^3+(1-p)^3$ and $c=[p^2+(1-p)^2]^2$, the anti-flatness maximum as a
function of the accumulated crossing number is
\begin{equation}
    m^*_{\cal F}(p)
    =
    \frac{\log\!\left(|\log c|/|\log b|\right)}
         {\log(b/c)} .
    \label{eq:supp-transport-mstar}
\end{equation}
This peak is controlled by the accumulated number of transported
non-flat dimers, whereas the entropy-growth-rate peak is controlled by
the transport flux $\dot m(t)$.  The two peak times therefore naturally
separate even when pure SWAP dynamics makes the transported non-flat
spectrum visible in ${\cal F}_A(t)$ for $p\neq1/2$.

\subsection{Localized dynamics and l-bit intuition}
\label{subsec:supp-lbit}

The MBL regime should not be identified with literal SWAP transport.
Rather, it resembles transport in the spectral sense used in the main
text: entropy can grow through the redistribution of phase correlations
across the bipartition without rapid local thermal scrambling of the
Schmidt weights.  A standard phenomenological description is the
$l$-bit Hamiltonian~\cite{PhysRevLett.110.260601,ROS2015420,Imbrie2016,PhysRevB.88.014206,RevModPhys.91.021001}
\begin{equation}
    H_{\rm lbit}
    =
    \sum_i h_i\tau_i^z
    +
    \sum_{i<j}J_{ij}\tau_i^z\tau_j^z
    +
    \sum_{i<j<k}J_{ijk}\tau_i^z\tau_j^z\tau_k^z+\cdots ,
    \label{eq:supp-lbit-h}
\end{equation}
with typical couplings decaying exponentially with range,
\begin{equation}
    |J_{ij}|\sim J_0 e^{-|i-j|/\xi(W)} .
    \label{eq:supp-lbit-decay}
\end{equation}
Starting from a product state in a physical basis, the state is
generically a coherent superposition in the $l$-bit basis.  Entanglement
then grows through dephasing between $l$-bits on opposite sides of the
cut.  This dephasing can broaden the active Schmidt support, but it does
not rapidly generate locally thermal, random-matrix-like Schmidt
sectors.  In this sense, localized entropy growth is closer to
transport-like redistribution than to local build.

Different spatial ranges activate at parametrically different dephasing
times,
\begin{equation}
    \tau_\ell(W)\sim |J_\ell(W)|^{-1},
    \qquad
    |J_\ell(W)|\sim J_0 e^{-\ell/\xi(W)} ,
    \label{eq:supp-lbit-times}
\end{equation}
so the entropy growth may be viewed schematically as a sum over
dephasing channels,
\begin{equation}
    S_A(t)
    \simeq
    S_{\rm loc}(t)
    +
    \sum_{\ell>1}
    s_\ell(W)
    f\!\left(\frac{t}{\tau_\ell(W)}\right).
    \label{eq:supp-lbit-rate-picture}
\end{equation}
Increasing disorder reduces $\xi(W)$ and suppresses the delayed,
longer-range channels.  The finite-time maximum of $\dot S_A(t)$ can
therefore become increasingly front-loaded: it is dominated by the
earliest local dephasing channel, while the later logarithmic tail
carries a weaker instantaneous rate.  This does not mean that stronger
disorder makes the full entanglement process faster; it means that the
later entropy-producing channels are suppressed, so the largest resolved
rate is pushed toward the shortest-range process.

The response of ${\cal F}_A$ is different.  Since
${\cal F}_A=P_3-P_2^2$ weights the largest Schmidt eigenvalues, it is
comparatively insensitive to the many tiny weights generated by weak,
distant dephasing channels.  The first anti-flatness peak is instead
tied to the earliest appreciable non-flat deformation of the dominant
Schmidt weights, and its height is reduced as local scrambling is
suppressed.  Localized dynamics therefore separates two clocks: the
maximum finite-time entropy-growth rate is controlled by the earliest
dephasing channel, whereas the magic barrier waits for non-flat spectral
deformation of the dominant Schmidt weights.  This naturally accounts
for the observed ordering $t_{\cal F}^*>t_{\dot S}^*$ and is consistent
with the growing peak separation $\Delta t_{\rm sep}$ with increasing
disorder.

\subsection{Peak-time trends from source--dilution competition}
\label{subsec:supp-peak-trends}

The peak-time trends observed in both the random-field XXZ chain
(Fig.~\ref{fig:peaktime}) and the SWAP--Haar circuit
(Fig.~3 of the main text and
Fig.~\ref{fig:supp-random-circuit-p07}) follow from the same
source--dilution logic in Eq.~\eqref{eq:supp-source-dilution}.  The
magic barrier appears when spectral roughening can no longer compensate
the dilution of low moments caused by the expanding Schmidt support.
Thus $t_{\cal F}^*$ is mainly set by the rate at which the dominant
Schmidt weights become nonuniform, whereas $t_{\dot S}^*$ is mainly set
by the fastest entropy-producing channel.

In the random-field XXZ chain, increasing $W$ suppresses delayed,
longer-range dephasing channels.  For both product and Bell-pair initial
states, this pushes the maximum of $\dot S_A(t)$ toward the earliest
short-range process, so $t_{\dot S}^*$ moves to earlier times without
implying faster overall entanglement growth.  The response of
$t_{\cal F}^*$ depends on the initial spectral content.  For product
states, the leading non-flat deformation of the Schmidt spectrum is
generated by early local dynamics and its timescale changes only weakly
with disorder.  For Bell-pair initial states, the transported Schmidt
blocks are initially flat; a nonzero magic barrier therefore requires
residual local dynamics to deform these blocks nonuniformly.  Stronger
disorder weakens this deformation channel, pushing $t_{\cal F}^*$ to
later times and producing the pronounced growth of
$\Delta t_{\rm sep}$.

The SWAP--Haar circuit provides a controlled counterpart.  At small
Haar fraction $r$, SWAP gates rapidly transport stored Bell-pair
entanglement, so $\dot S_A(t)$ peaks at the earliest resolved times,
while the magic barrier waits for sparse Haar gates to deform the flat
transported Schmidt blocks.  Increasing $r$ directly strengthens this
local roughening source, pulls $t_{\cal F}^*$ toward the early
entropy-growth window, and decreases $\Delta t_{\rm sep}$.  The circuit
and XXZ trends are therefore two manifestations of the same
source--dilution logic: the peak separation diagnoses whether entropy
growth is accompanied immediately by local spectral reshaping, or first
proceeds through transport-like redistribution of pre-existing
entanglement.

\section{Haar-state benchmark for anti-flatness}
\label{sec:supp-haar-benchmark}

In this section we provide the anti-flatness of Haar random pure states~\cite{PhysRevLett.71.1291,10.1063/1.523763,KarolZyczkowski_2001,Hans-JurgenSommers_2004}.
This provides a useful flat-spectrum reference for the small
anti-flatness expected in thermalizing systems.

\begin{figure}[t]
    \centering
    \includegraphics[width=0.85\textwidth]{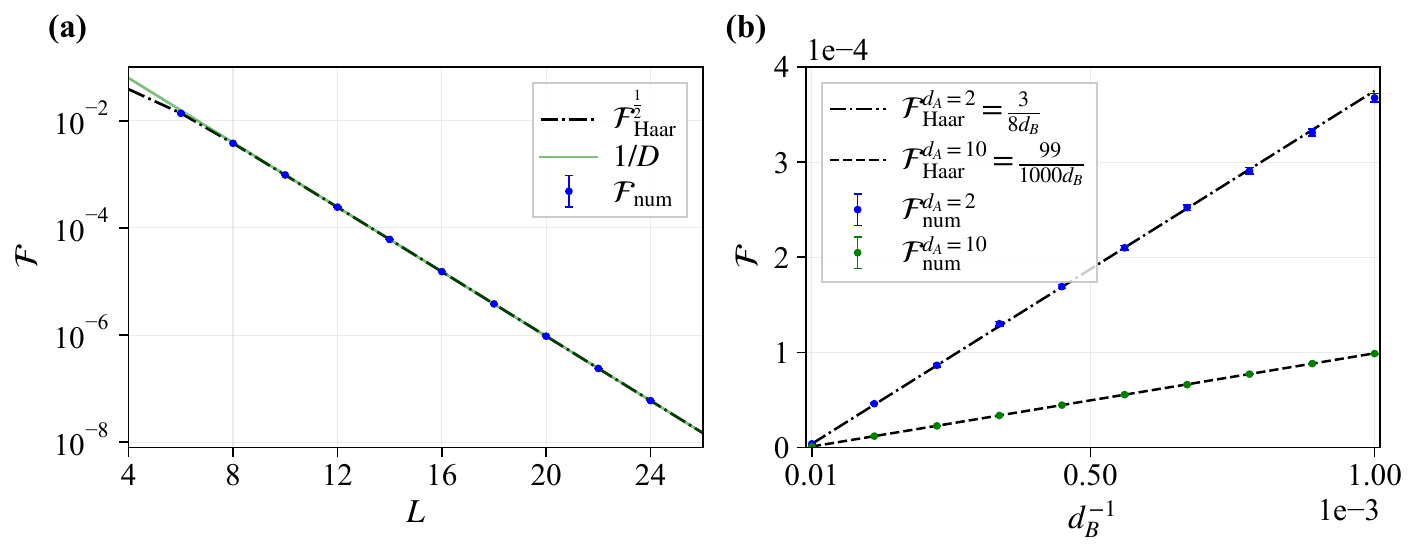}
    \caption{
    Haar anti-flatness benchmarks.  (a) Equal-bipartition anti-flatness
    as a function of spin-$1/2$ chain length $L$.  The black
    dash-dotted curve is Eq.~\eqref{eq:supp-haar-half}, the green line
    shows the leading $1/D$ scaling, and the blue points are numerical
    Haar averages.  (b) Fixed-subsystem anti-flatness for $d_A=2$ and
    $d_A=10$ as a function of $d_B^{-1}$; dashed curves are
    Eq.~\eqref{eq:supp-haar-fixedA}.}
    \label{HaarAF}
\end{figure}

Consider a bipartite Hilbert space with subsystem dimensions $d_A$ and
$d_B$, and total dimension $D=d_A d_B$.  The Haar-averaged
anti-flatness is
\begin{equation}
    {\cal F}_{\rm Haar}
    =
    \mathbb{E}_{\psi}\!\left[\operatorname{Tr}\rho_A^3\right]
    -
    \mathbb{E}_{\psi}\!\left[
        \left(\operatorname{Tr}\rho_A^2\right)^2
    \right] .
    \label{eq:supp-haar-def}
\end{equation}
Using standard Haar moment formulas~\cite{Mele2024introductiontohaar},
\begin{align}
    \mathbb{E}_{\psi}\!\left[\operatorname{Tr}\rho_A^3\right]
    &=
    \frac{d_A^2+3d_A d_B+d_B^2+1}
    {(d_A d_B+1)(d_A d_B+2)}, \\
    \mathbb{E}_{\psi}\!\left[
        \left(\operatorname{Tr}\rho_A^2\right)^2
    \right]
    &=
    \frac{
    d_A^3d_B+d_A d_B^3+2d_A^2d_B^2+10d_A d_B
    +4d_A^2+4d_B^2+2}
    {(d_A d_B+1)(d_A d_B+2)(d_A d_B+3)} .
    \label{eq:supp-haar-moments}
\end{align}
For an equal bipartition, $d_A=d_B=\sqrt D$, this gives
\begin{equation}
    {\cal F}^{1/2}_{\rm Haar}
    =
    \frac{(D-1)^2}{(D+1)(D+2)(D+3)}
    =
    \frac{1}{D}-\frac{8}{D^2}
    +O(D^{-3}) .
    \label{eq:supp-haar-half}
\end{equation}
Thus the equal-bipartition Haar anti-flatness vanishes as the total
Hilbert-space dimension increases.  For fixed $d_A$ and large $d_B$,
\begin{equation}
    {\cal F}_{\rm Haar}^{d_A}
    =
    \frac{d_A^2-1}{d_A^3}\frac{1}{d_B}
    +O(d_B^{-2}) .
    \label{eq:supp-haar-fixedA}
\end{equation}
The smallness of Haar anti-flatness reflects the fact that the reduced
density matrix is close to the normalized identity on a large support.
Figure~\ref{HaarAF} verifies these scalings numerically.

\section{Disordered XXZ chain: nonequilibrium dynamics, peak extraction, and stationary limits}
\label{sec:supp-xxz-framework}

This section provides additional numerical details for the random-field
spin-$1/2$ XXZ chain studied in the main text.  We first present
representative nonequilibrium metrics and specify the peak-extraction
convention.  We then give two complementary references for the
Schmidt-spectrum shape: the anti-flatness of many-body eigenstates and
the long-time plateau reached after quench dynamics.

\subsection{Dynamical traces and peak-time extraction}
\label{subsec:supp-peak-convention}

We consider the random-field spin-$1/2$ XXZ chain governed by the
Hamiltonian defined in the main text,
\begin{equation}
    H
    =
    J\sum_{i=1}^{L}
    \left(
        \sigma_i^x\sigma_{i+1}^x
        +
        \sigma_i^y\sigma_{i+1}^y
    \right)
    +
    \Delta\sum_{i=1}^{L}\sigma_i^z\sigma_{i+1}^z
    +
    \sum_{i=1}^{L}h_i\sigma_i^z ,
    \label{eq:supp-xxz-hamiltonian}
\end{equation}
with periodic boundary conditions.  We set $J=\Delta=1$ and sample the
random longitudinal fields independently from $h_i\in[-W,W]$.  The
dynamics is computed in the half-filling sector,
$\sum_i\sigma_i^z=0$.

\begin{figure}[tb]
    \centering
    \includegraphics[width=0.85\textwidth]{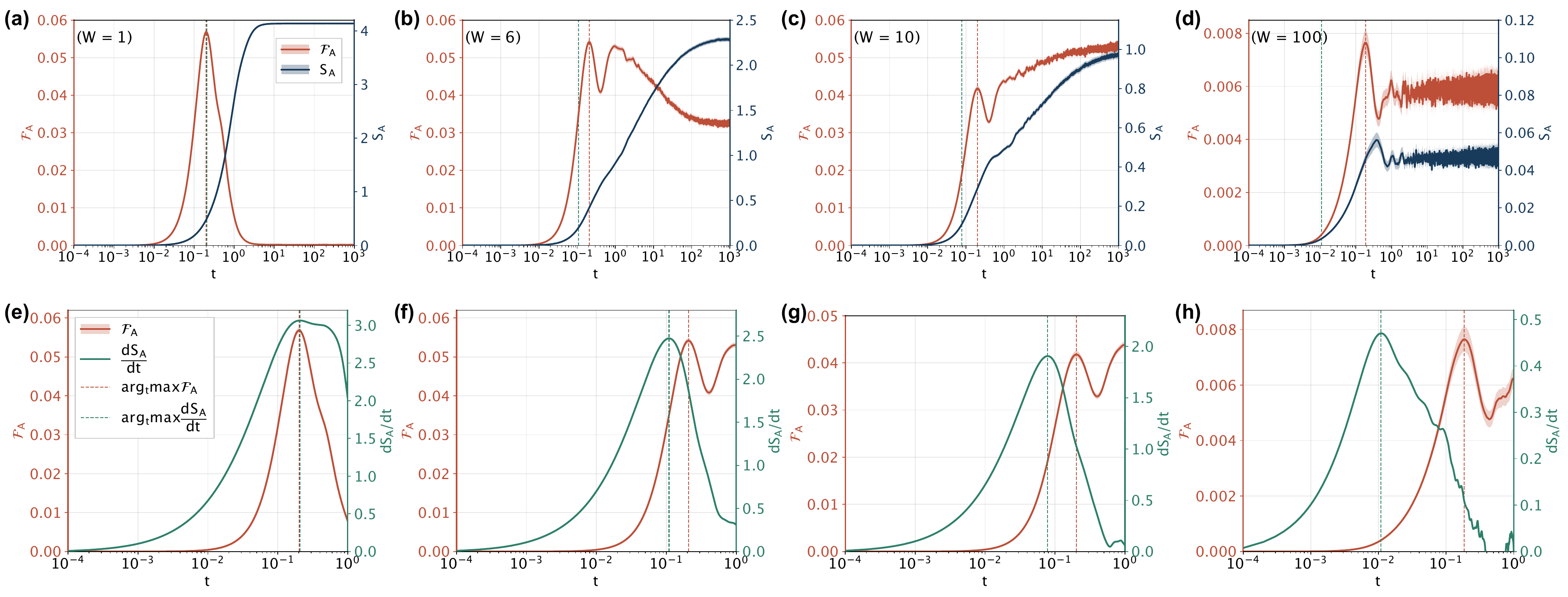}
    \caption{
    Product-state dynamics in the random-field XXZ chain.  The panels
    show representative ensemble-averaged traces of ${\cal F}_A(t)$,
    $S_A(t)$, and $\dot S_A(t)$ for different disorder strengths.  The
    diagnostic used in the main text compares the first transient
    anti-flatness maximum with the first maximum of the entropy-growth
    rate.}
    \label{AFDynamicProduct}
\end{figure}

Figure~\ref{AFDynamicProduct} shows representative product-state quench
dynamics.  At weak disorder, ${\cal F}_A(t)$ develops a transient
barrier in the same early-time window in which the entropy-growth rate
$\dot S_A(t)$ is largest, consistent with local build-dominated
dynamics.  As $W$ is increased, the first anti-flatness peak remains tied
to a microscopic local spectral response, while the first maximum of
$\dot S_A(t)$ shifts toward earlier times.  Near the thermal--MBL
crossover, later-time features of ${\cal F}_A(t)$ can become comparable
to the first barrier.  To isolate the short-time spectral response that
is compared with entropy growth in the main text, we therefore use the following
first-peak convention. For each disorder strength, we first average $S_A(t)$ and
${\cal F}_A(t)$ over disorder realizations and initial states.  The
growth rate $\dot S_A(t)$ is extracted from the ensemble-averaged
entropy trace by finite differences.  We define
\begin{equation}
    t_{\cal F}^*
    =
    \text{first local-maximum time of }{\cal F}_A(t),
    \qquad
    t_{\dot S}^*
    =
    \text{first local-maximum time of }\dot S_A(t),
    \label{eq:supp-peak-times}
\end{equation}
and
\begin{equation}
    \Delta t_{\rm sep}=t_{\cal F}^*-t_{\dot S}^* .
    \label{eq:supp-peak-separation}
\end{equation}
For Bell-pair initial states
at very strong disorder, the first peak of $\dot S_A(t)$ is broadened by
ensemble averaging; the peak-time analysis is therefore restricted to
the parameter regime where both peak positions can be identified
stably.

\begin{figure}[tb]
    \centering
    \includegraphics[width=0.85\textwidth]{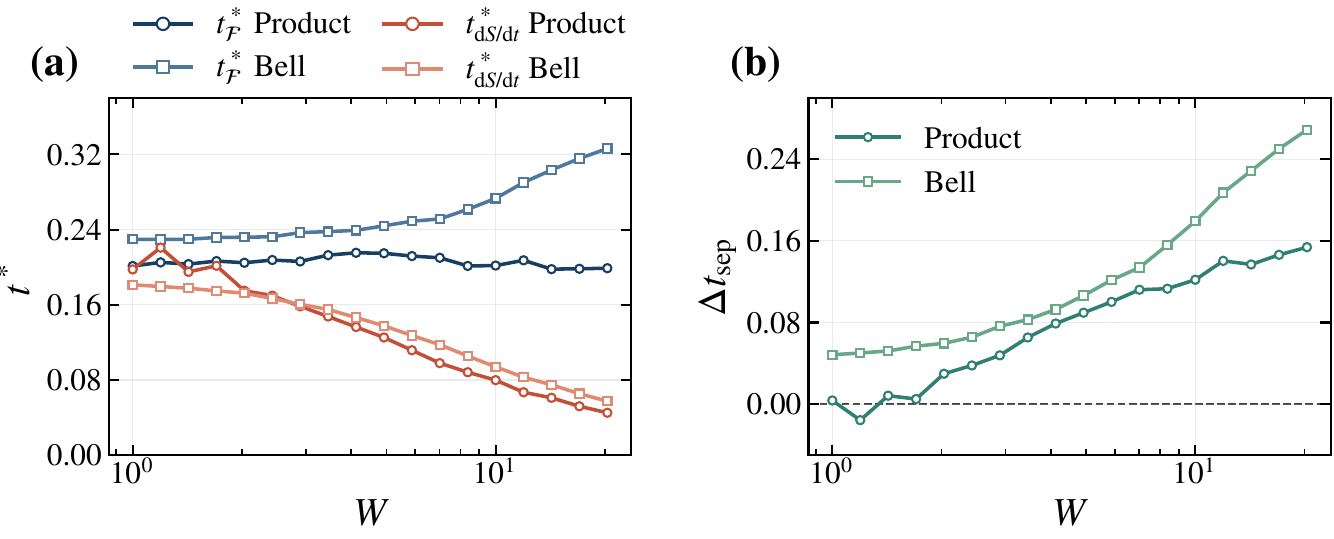}
    \caption{
    Peak times and peak separation in the random-field XXZ chain at
    $L=14$.  (a) The first anti-flatness peak time $t_{\cal F}^*$ and
    the first entropy-growth-rate peak time $t_{\dot S}^*$ for product
    and Bell-pair initial states.  (b) The corresponding separation
    $\Delta t_{\rm sep}=t_{\cal F}^*-t_{\dot S}^*$.  Each data point is
    averaged over at least $1000$ disorder realizations.}
    \label{fig:peaktime}
\end{figure}

The extracted peak positions and the resulting separation are shown in
Fig.~\ref{fig:peaktime}.  For product initial states, $t_{\cal F}^*$ is
only weakly dependent on $W$ over a broad range, whereas
$t_{\dot S}^*$ moves toward earlier times as the disorder strength $W$ increases.  For Bell-pair initial states, the separation is
enhanced in the localized regime, consistent with stored, nearly flat
entanglement being redistributed across the half-chain bipartition
before
substantial nonuniform spectral deformation develops.  The separation in
Fig.~\ref{fig:peaktime}(b) corresponds to the data plotted in
Fig.~2(e) of the main text.

\begin{figure}[tb]
    \centering
    \includegraphics[width=0.5\textwidth]{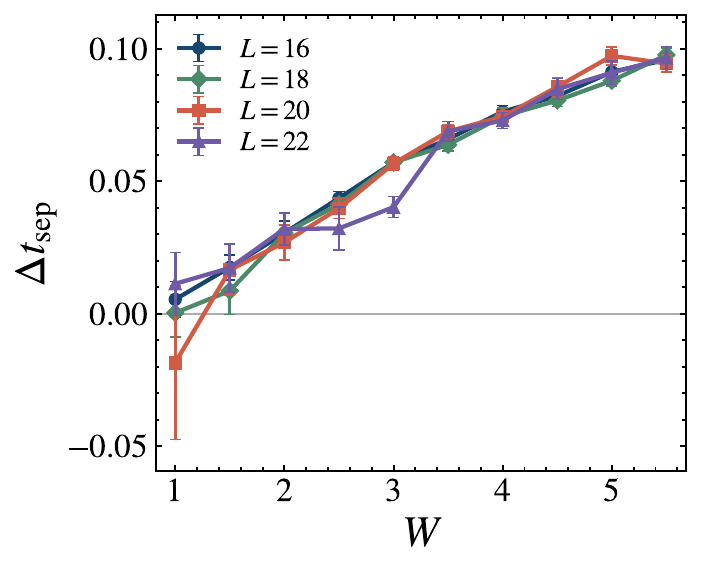}
    \caption{Finite-size stability of the relative peak time in the disordered XXZ chain.
We show the peak-time separation
\(\Delta t_{\rm sep}=t_{\cal F}^*-t_{\dot S}^*\) as a function of disorder strength
\(W\) for \(L=16,18,20,22\).
The subsystem \(A\) is half of the chain, and the initial states are random
product states in the half-filling sector.  The dynamics are computed using
sparse Krylov real-time evolution.  The near collapse of different system sizes
shows that the relative delay of the anti-flatness peak is stable over the
accessible sizes.}
    \label{fig:peak-separation-size}
\end{figure}

To test the finite-size stability of the relative peak times, we further compute
\(S_A(t)\) and \(F_A(t)\) for larger chains using sparse Krylov real-time
evolution~\cite{Zhang2023tensorcircuit,zhang2026tensorcircuitnguniversalcomposablescalable}.
The peak-time separation is shown in Fig.~\ref{fig:peak-separation-size}.  For \(L=16,18,20,22\),
\(\Delta t_{\rm sep}\) is essentially size independent within statistical
uncertainties, indicating that the observed
relative delay of the anti-flatness peak is not a small-system artifact.

\subsection{Eigenstate anti-flatness}
\label{subsec:supp-eigen-af}

As a complementary static reference for the entanglement-spectrum shape,
we compute anti-flatness in many-body eigenstates $|E_n\rangle$ of the
disordered XXZ Hamiltonian,
\begin{equation}
    \rho_A^{(n)}=\operatorname{Tr}_B|E_n\rangle\langle E_n|,
    \qquad
    {\cal F}_A^{(n)}
    =
    \operatorname{Tr}\left[(\rho_A^{(n)})^3\right]
    -
    \left\{\operatorname{Tr}\left[(\rho_A^{(n)})^2\right]\right\}^2 .
    \label{eq:supp-eigen-af}
\end{equation}
The calculation is performed in the half-filling sector for
$L=6,\ldots,16$.  For $L<14$, we average over the full spectrum; for
$L\ge14$, we average over the $50$ eigenstates closest to the middle of
the spectrum.  Each data point is averaged over at least $2500$ disorder
realizations.

\begin{figure}[tb]
    \centering
    \includegraphics[width=0.5\textwidth]{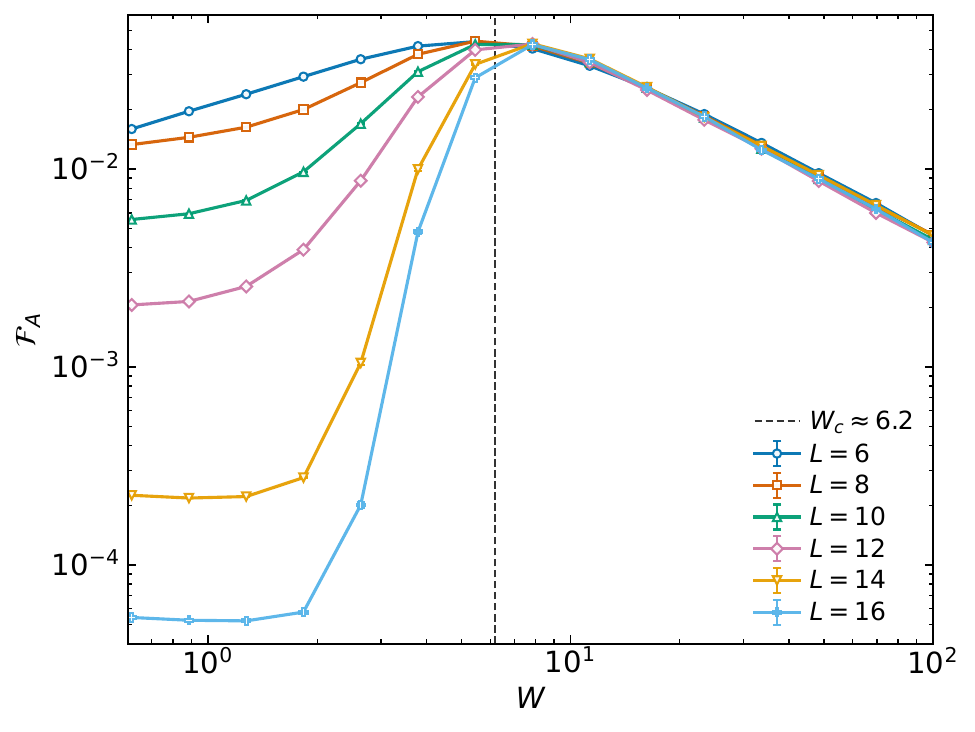}
    \caption{
    Eigenstate anti-flatness in the random-field XXZ chain.  The
    half-chain anti-flatness ${\cal F}_A^{(n)}$ is averaged over
    eigenstates in the half-filling sector and over disorder
    realizations.}
    \label{EigenAFXXZ}
\end{figure}

As shown in Fig.~\ref{EigenAFXXZ}, eigenstate anti-flatness is small
deep in the thermal regime, becomes enhanced near the thermal--MBL
crossover, and is suppressed again at strong disorder.  This
nonmonotonic behavior is consistent with anti-flatness being largest
when the Schmidt spectrum is neither Haar-flat on a large support nor
close to a weakly entangled product-like structure.

\subsection{Long-time anti-flatness after a quench}
\label{subsec:supp-long-time-af}

Finally, we examine the anti-flatness plateau reached at long times
after a quench.  We define
\begin{equation}
    {\cal F}_A^{\infty}(W,L)
    =
    \lim_{T\to\infty}
    \frac{1}{T-t_0}
    \int_{t_0}^{T} \mathrm{d}\tau\,
    \left\langle{\cal F}_A(\tau)\right\rangle ,
    \label{eq:supp-late-af}
\end{equation}
where $\langle\cdots\rangle$ denotes averaging over disorder
realizations and initial states.  Numerically, we set $t_0=10^4$ and
average over a late-time window after relaxation to a plateau, using at
least $500$ disorder realizations.

\begin{figure}[tb]
    \centering
    \includegraphics[width=1\textwidth]{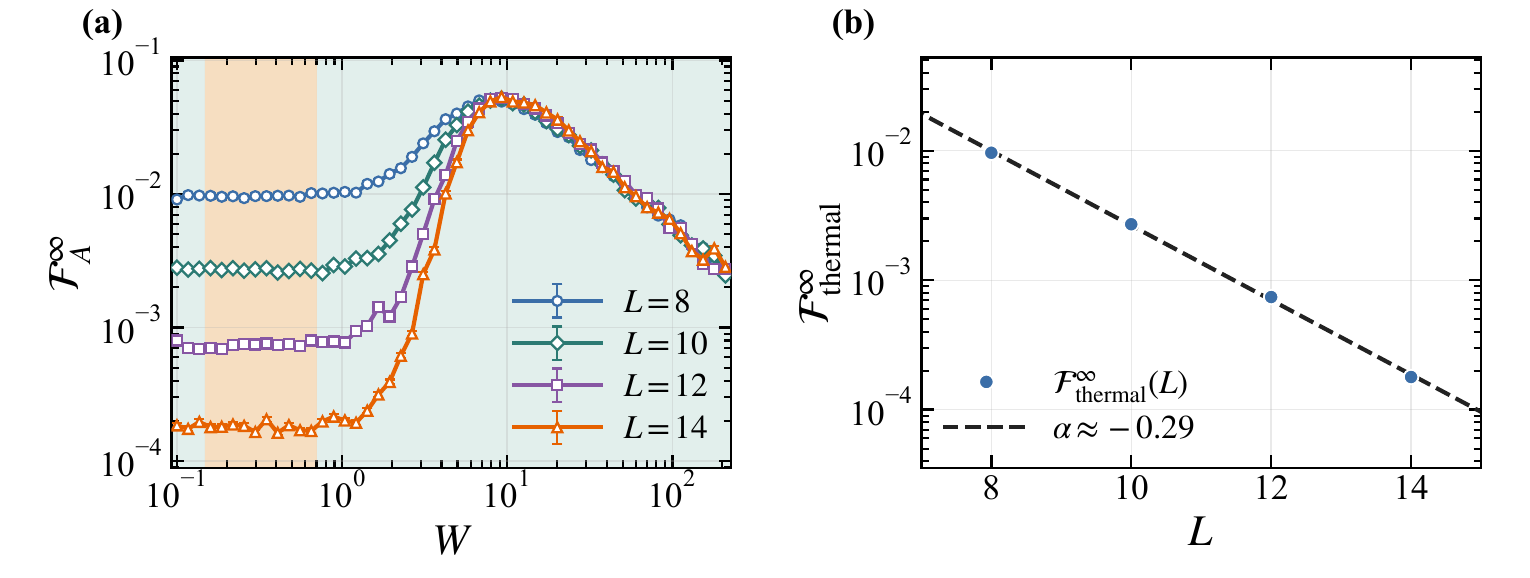}
    \caption{
    Long-time anti-flatness in the random-field XXZ chain.  (a)
    Long-time-averaged half-chain anti-flatness as a function of disorder
    strength $W$ for different system sizes. (b) Thermal-side finite-size scaling obtained by averaging the data in the orange-shaded region of panel (a), compared with an
    exponential fit close to the Haar expectation. }
    \label{fig:XXZLongF}
\end{figure}

As shown in Fig.~\ref{fig:XXZLongF}(a), the long-time anti-flatness
${\cal F}^{\infty}$ is nonmonotonic in $W$, similar to the eigenstate
trend in Fig.~\ref{EigenAFXXZ}.  In the thermal regime, the plateau
value decreases approximately as $2^{-L}$ [Fig.~\ref{fig:XXZLongF}(b)],
consistent with a large nearly flat thermal support.

\section{Random-circuit benchmark with stored entanglement}
\label{sec:supp-random-circuit-benchmark}

In the main text, we use a tunable random circuit to isolate transport-
and build-dominated contributions to entanglement growth.  Here we
provide the circuit protocol and additional data for partially entangled
initial dimers.

\begin{figure}[t]
    \centering
    \includegraphics[width=0.6\textwidth]{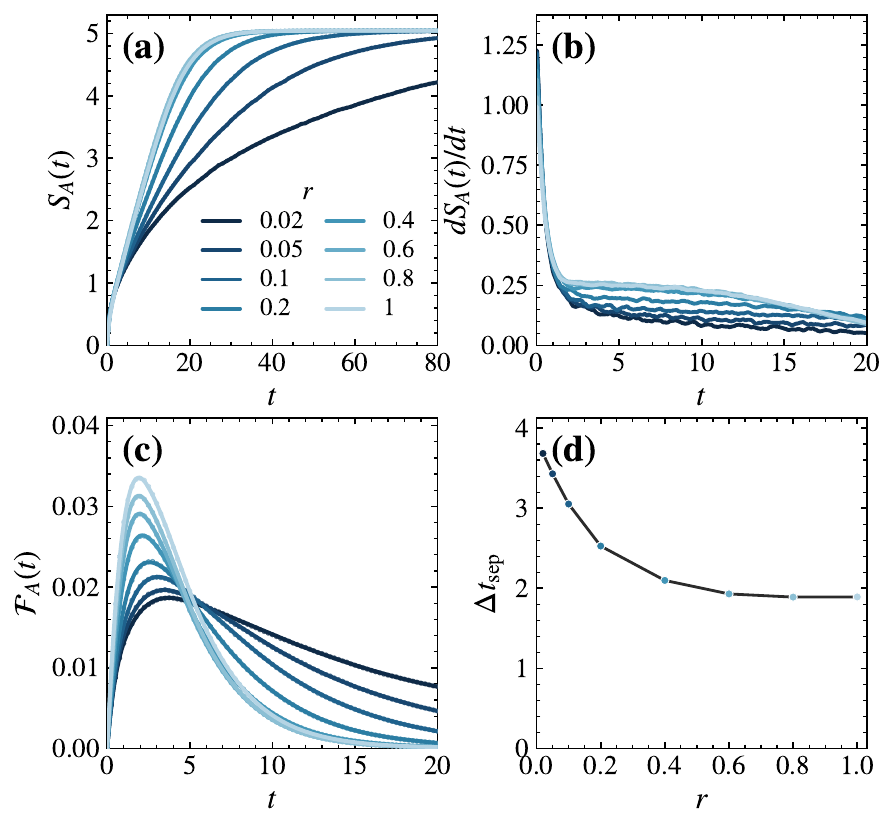}
    \caption{
    Random-bond circuit initialized from partially entangled dimers with
    $p=0.7$.  The protocol and statistics are the same as in the
    Bell-pair benchmark shown in Fig. 3 of the main text.
    Because transported dimers already have a non-flat two-level Schmidt
    spectrum, transport can contribute directly to ${\cal F}_A(t)$.
    Nevertheless, increasing the Haar-gate fraction $r$ pulls the
    anti-flatness peak toward the early entropy-growth window.}
    \label{fig:supp-random-circuit-p07}
\end{figure}

We consider an open spin-$1/2$ chain of length $L$, bipartitioned at
the center into
\begin{equation}
    A=\{1,\ldots,L/2\},
    \qquad
    B=\{L/2+1,\ldots,L\}.
    \label{eq:supp-random-circuit-cut}
\end{equation}
The initial state stores entanglement inside each half-chain, with no
dimer crossing the central cut,
\begin{equation}
    |\psi_0(p)\rangle
    =
    \prod_{(i,j)\in{\cal D}_A\cup{\cal D}_B}
    |\phi_p\rangle_{ij},
    \qquad
    |\phi_p\rangle_{ij}
    =
    \sqrt p\,|01\rangle_{ij}
    +
    \sqrt{1-p}\,|10\rangle_{ij}.
    \label{eq:supp-random-circuit-initial-state}
\end{equation}
This family interpolates between a product state at $p=0$ or $p=1$ and
a Bell-pair state at $p=1/2$.  For the data shown in the main text and
in Fig.~\ref{fig:supp-random-circuit-p07}, we take $L=16$ and choose
\begin{equation}
    {\cal D}_A=\{(1,2),(3,4),(5,6),(7,8)\},
    \qquad
    {\cal D}_B=\{(9,10),(11,12),(13,14),(15,16)\}.
    \label{eq:supp-random-circuit-dimers}
\end{equation}
Thus $S_A(0)=0$, although each half-chain contains a reservoir of stored
short-range entanglement.

The endpoint $p=1$ is not a useful benchmark for transport.  In this
case $|\phi_{p=1}\rangle=|01\rangle$, so the initial state is a product
state on individual sites and contains no pre-existing entanglement that
SWAP gates can move into the bipartition.  We therefore focus on
$p=1/2$, where transported Bell pairs remain spectrally flat, and on
$p=0.7$, where a transported dimer already carries a controlled non-flat
two-level Schmidt spectrum.

The dynamics is generated by random nearest-neighbor updates.  At each
elementary update, a bond $(i,i+1)$ is chosen uniformly at random and
acted on by
\begin{equation}
    U_{i,i+1}
    =
    \begin{cases}
        \mathrm{SWAP}_{i,i+1}, & \text{with probability }1-r,\\
        U_{\rm Haar}\in U(4), & \text{with probability }r ,
    \end{cases}
    \label{eq:supp-random-circuit-update}
\end{equation}
where $U_{\rm Haar}$ is a Haar random two-qubit gate.  One sweep
consists of $L$ elementary updates and corresponds to $\Delta t=1$.
The parameter $r$ is the build fraction: SWAP gates move stored
entanglement without locally creating entanglement, whereas Haar random
gates locally build and scramble entanglement.

The data in Fig.~\ref{fig:random-circuit-p05} of the main text and in
Fig.~\ref{fig:supp-random-circuit-p07} are averaged over
$N_{\rm traj}=50000$ independent circuit realizations.  Observables are
sampled with time step $dt=1/L=1/16$ for $t\le20$ and with $dt=1$ for
$t>20$.  The growth rate $\dot S_A(t)$ is computed by first smoothing
the ensemble-averaged entropy trace and then taking a finite difference.
For ${\cal F}_A(t)$, symbols denote representative ensemble means with
standard errors, while solid curves are smoothing splines fitted to the
high-resolution time series and used for peak extraction.  The peak
separation $\Delta t_{\rm sep}$ is defined as in Eq.~\eqref{eq:tsep} of
the main text.

The Bell-pair initial state, $p=1/2$, gives the cleanest separation
between transport and local build.  In the pure transport limit,
moving intact Bell pairs across the central cut increases $S_A(t)$ but
leaves the entanglement spectrum flat, so ${\cal F}_A=0$.  Consequently, the early maximum of
$\dot S_A(t)$ in Fig.~\ref{fig:random-circuit-p05} is driven by the
rapid motion of stored entanglement.  By contrast, the delayed maximum
of ${\cal F}_A(t)$ requires Haar gates to nonuniformly deform the flat
Bell-pair Schmidt blocks.  Increasing $r$ strengthens this local build
channel and reduces $\Delta t_{\rm sep}$.

The $p=0.7$ data provide an intermediate check.  Since a transported
dimer is already non-flat, the transient growth of ${\cal F}_A(t)$ is no
longer purely build-generated.  Nevertheless, the same trend persists:
transport-dominated circuits exhibit a larger delay between entropy
growth and spectral deformation, whereas increasing the build fraction
pulls the two maxima into the same early-time window.  This controlled
circuit benchmark is consistent with the XXZ results: the peak
separation is minimized when entropy growth is accompanied by local
spectral reshaping.

\end{document}